\documentclass[a4paper,10pt,twoside]{cpc-hepnp}

\usepackage{multicol}
\usepackage{graphicx}
\usepackage{booktabs}
\usepackage{amssymb,bm,mathrsfs,bbm,amscd}
\usepackage[tbtags]{amsmath}
\usepackage{lastpage}

\begin{document}
%\begin{CJK*}{GBK}{song}

\fancyhead[c]{\small Submitted to Chinese Physics C} \fancyfoot[C]{\small 010201-\thepage}
%\fancyhead[c]{\small Chinese Physics C~~~Vol. XX, No. X (201X)
%XXXXXX} \fancyfoot[C]{\small 010201-\thepage}

\footnotetext[0]{Received 14 March 2009}

\title{The influence of experimental setup on the spectroscopy investigation of $^{\mathrm{14}}$Be by Coulomb breakup reaction  \thanks{Supported by National Natural Science
Foundation of China (11205036) }}
\author{%
      SONG Yu-Shou$^{1;1)}$\email{songyushou80@163.com}%
\quad HOU Ying-Wei$^{1}$%
\quad HU Li-Yuan$^{1}$
\quad LIU Hui-Lan$^{1}$
\quad WU Hong-Yi$^{1}$
}
\maketitle
\address{%
$^1$ Key Discipline Laboratory of Nuclear Safety and Simulation Technology, Harbin Engineering University, Harbin 150001, China\\
%$^2$ State key laboratory of nuclear physics and technology, Peking University, Beijing 100187, China\\
%$^3$ Institute of Modern Physics, Chinese Academy of Sciences, Lanzhou 730000, China\\
%$^4$ College of Nuclear Engineering, Chongqing University, Chongqing 400044, China\\
}
\begin{abstract}
The two-body core+$2n$ cluster structure was implemented to describe the two-neutron halo nucleus $^{\mathrm{14}}\mathrm{Be}$, where the core$^{\mathrm{12}}\mathrm{Be}$ was assumed inert and at ground state and the dineutron was assumed at pure $2S_0$ state. Based on such a structure the three-body continuum-discretized coupled-channel (CDCC) calculation was successfully used to deal with the $^{\mathrm{14}}\mathrm{Be}$ breakup reactions of $^{\mathrm{14}}\mathrm{Be}+^{\mathrm{12}}\mathrm{C}$ at 68~MeV/nucleon and $^{\mathrm{14}}\mathrm{Be}+ $Pb at 35~MeV/nucleon.
%Through the exist $^{\mathrm{14}}\mathrm{Be}+^{\mathrm{12}}\mathrm{C}$ experiment data the normalizing factor for the relative energy spectrum was extracted as 0.27. With this normalizing factor, the CDCC calculation was also used to solve the Coulomb breakup of $^{\mathrm{14}}\mathrm{Be}$ at a lead target. The theoretical result of energy spectroscopy agrees with the experimental data given by M. Labiche et al. at GANIL.
%In ion-ion reactions at energies of tens to hundred MeV/nucleon, the fraction of peripheral collisions reaches around 20\% $\sim$ 30\%. This makes Coulomb breakup become a powerful tool to investigate the neutron-halo nuclei especially those with low beam intensity provided by the radioactive ion beam (RIB) facility.
%The three-body continuum-discretized coupled-channel (CDCC) method was used for for $^{\mathrm{14}}\mathrm{Be}$ breakup reaction, where the nucleus $^{\mathrm{14}}\mathrm{Be}$ was
%treated as core+dineutron. Based on the $^{\mathrm{14}}\mathrm{Be}+\mathrm{12}C$ experiment data from reference the spectroscopic factor of the dineutron was extracted as 0.27. The theoretical results of Coulomb breakup reaction of $^{\mathrm{14}}\mathrm{Be}$ are in consent with the experiment.
Consequently, we modeled the kinematically complete measurement experiment of $^{\mathrm{14}}\mathrm{Be}$ (35~MeV/nucleon) Coulomb breakup at a lead target with the help of Geant4. From the simulation data the relative energy spectrum was constructed by the invariant mass method and $B(E1)$ spectrum was extracted using virtual photon model. The influence of the target thickness and detector performance on the spectroscopy was investigated.

\end{abstract}

\begin{keyword}
Coulomb breakup, halo nucleus, CDCC, dineutron
\end{keyword}

\begin{pacs}
25.70.De %Coulomb excitation
29.30 %Spectrometers and spectroscopic techniques
\end{pacs}

\footnotetext[0]{\hspace*{-3mm}\raisebox{0.3ex}{$\scriptstyle\copyright$}2013
Chinese Physical Society and the Institute of High Energy Physics
of the Chinese Academy of Sciences and the Institute
of Modern Physics of the Chinese Academy of Sciences and IOP Publishing Ltd}%

\begin{multicols}{2}
\section{Introduction}
%$^{\mathrm{11}}$Li $^{\mathrm{9}}$Li
Since the neutron halo structure was discovered in 1985~\cite{prl55.2676}, it has attracted so much interests of the nuclear physicists in the world. Besides the large nuclear matter radius and the narrow transverse momentum distribution of the core~\cite{prl69.2050}, the anomalously large Coulomb breakup cross section is another halo manifestation. With the radioactive ion beam (RIB) facility upgrade and experimental techniques development, the kinetically complete measurement of the Coulomb breakup together with the invariant mass spectroscopy have become a standard research approach of halo nucleus properties. Through the investigations of one-neutron halo nuclei like $^{\mathrm{11}}$Be~\cite{prc70.054606, prc68.034318}, $^{\mathrm{15}}$C~\cite{prc79.035805}, it has been found that the cross section enhancement is due to the spatially decouple of the halo relative to the core. A direct breakup mechanism was concluded according to the $B(E1)$ spectrum analysis. There are similar soft Coulomb excitation for two-neutron halo nuclei, such as $^{\mathrm{6}}$He~\cite{prc65.034306}, $^{\mathrm{11}}$Li~\cite{prl96.252502} and $^{\mathrm{14}}$Be~\cite{prl86.600}. The understanding of the nature of the $E1$ excitation for two-neutron halo nuclei depends on the experimental spectroscopy. However, the experimental work is more complicated and difficult to carry out. The inefficiency of correlated two-neutron detection and finitely thin target approximation may lead to distorted energy spectroscopy, which had been indicated in the $^{\mathrm{11}}$Li~\cite{prl96.252502} relative spectrum measurement.

In order to discover how the experimental setting influence the energy spectroscopy of $^{\mathrm{14}}$Be Coulomb breakup on Pb target, the kinetically complete measuring experiment was modeled with the help of Geant4 Monte Carlo toolkit~\cite{IEEEns53.270}. The transportation of the projectile and outgoing particles were handled by Geant4 provided physical processes. To deal with the breakup of a two-halo, a so-called four-body CDCC, including a three-body projectile~\cite{plb252.311} and a target, has been developed independently by two groups~\cite{prc73.051602,prc77.064609}. However, such a calculation is still at a primary phase and principally achieved a success to calculate the reaction of $^{\mathrm{6}}$He up to now. According to the theoretical structure investigation of two-neutron halo nuclei, the dineutron correlation plays a dominate role~\cite{prc72.044321}.
Therefore, the Coulomb breakup of $^{\mathrm{14}}$Be was studied by the so-called three-body CDCC~\cite{pr154.125} calculation (FRESCO code), with a two-body projectile ($^{\mathrm{12}}$Be+dineutron) and a target nucleus. The theoretical results of energy spectrum and angular distribution of neutron agree with the exist experimental data. Based on the data output of the modeled experiment system, the relative kinetic energy spectrum was reconstructed by the invariant mass method, and consequently the $B(E1)$ spectrum was extracted by the virtual photon model of electromagnetic dissociation (EMD)~\cite{pr163.299}. The influence of the target thickness and the performances of detector system behind the target on the spectroscopy was discussed.

%
%Behind the target a charged ion telescope composed of a silicon strip detector and a CsI scintillator array was used to detector the charge products $^{\mathrm{12}}$Be, and a neutron wall consisting ?? pieces of plastic scintillator bars was responsible for the neutron detection.
%\section{Experiment modeling}%Approaches
\section{The Coulomb breakup of $^{\mathrm{14}}$Be+Pb}
%A 2n halo nucleus dissociates into $^{\mathrm{12}}$Be+$n$+$n$ when it bombards. %by soft $E1$ excitation
The breakup process of $^{\mathrm{14}}$Be on the high Z target~(Pb) consists two aspects~---~the structure part and reaction part.
%We utilized the CDCC method~\cite{pr154.125} to calculate the Coulomb breakup of $^{\mathrm{14}}$Be on a lead target.
As for the structure, $^{\mathrm{14}}$Be was treated a $^{\mathrm{12}}$Be core plus a dineutron (core + $2n$), which has a spin of zero~\cite{prc72.037603,prc76.044323,prc70.014603}. According to the Pauli principle $NL$ of the dineutron equals $2S$. Ignoring the internal motion of the dineutron, the binding energy between $^{\mathrm{12}}$Be and the dineutron is taken as the two-neutron separation energy $S_{2n}$ for the ground state. The distance between $^{\mathrm{12}}$Be and the dineutron was set as 5.6(9) fm according to the calculation with few-body reaction model for $^{\mathrm{14}}$Be~\cite{npa658.313}. Only the first $2^+$ resonance state at excitation energy 1.54~(13)~MeV, i.e., at 0.27~MeV above the breakup threshold was taken into account\cite{plb654.160}. %Assuming $^{\mathrm{14}}$Be and its core $^{\mathrm{12}}$Be lie on their ground states,
Assuming $^{\mathrm{14}}$Be and the core $^{\mathrm{12}}$Be both lie at ground state, the Woods-Saxon potential with the parameters listed in Table~\ref{tab1} was implemented to calculated their wave functions. The first $2^+$ state has the same potential parameters except the depth of the binding potential $V_0$, which was adjusted according to the excited energy.
%The spectroscopic factor the dineutron for s-wave is chosen as 0.27, which was validated by a comparison (Fig.~\ref{fig:spec}-a) between the CDCC calculation of $^{\mathrm{14}}$Be breakup on $^{\mathrm{12}}$C target and corresponding experimental result~\cite{plb654.160}.

\begin{center}
\tabcaption{\label{tab1}The Woods-Saxon potential parameters for ground state and first $2^+$ state of two-body $^{\mathrm{14}}$Be.}
\footnotesize
\begin{tabular}{lllllll}%{50mm}@{\extracolsep{\fill}}
\toprule
  &   $a_1$  &   $a_2$    & $r_c$ & $V_0$   & $r_0$ & $a_0$\\
  &          &            &  (fm) & (MeV)   &   (fm)& (fm)  \\
\hline
$^{\mathrm{14}}\mathrm{Be}_{g.s.}$  &0  &12  &1.2  &38.85 &1.81  &0.65\\
$^{\mathrm{14}}\mathrm{Be}_{2^+}$   &0  &12  &1.2  &35.28 &1.81  &0.65\\
\bottomrule
\end{tabular}
\end{center}
%=^{\mathrm{12}}\mathrm{Be}+2n=^{\mathrm{12}}\mathrm{Be}+2n

The so-called three-body CDCC method was implemented to deal with the breakup reaction of $^{\mathrm{14}}$Be. Using the nuclear structure described above, the $^{\mathrm{14}}\mathrm{Be}$ nuclear breakup reaction for $^{\mathrm{14}}\mathrm{Be}+^{\mathrm{12}}\mathrm{C}$ at 68~MeV/nucleon was calculated, where $s$, $p$, $d$ partial waves were considered. With a normalizing factor of 0.27 the calculated relative energy spectrum agree pretty well with that given by the experiment \cite{plb654.160} (Fig.~\ref{fig:spec}-a).
%According to the relative energy spectrum from the experiment \cite{plb654.160} a.
The CDCC method was also utilized to treat the $^{\mathrm{14}}$Be breakup at a Pb target. Referring to the one-neutron halo nuclei, the direct breakup was assumed for $^{\mathrm{14}}$Be. So that only s-wave and p-wave were considered, while d- and above order waves which including resonance states were ignored. As for the optical potential between the outgoing particles and the target nucleus, the potential between $^{\mathrm{12}}$Be and $^{\mathrm{208}}$Pb was substituted by $^{\mathrm{13}}$C+$^{\mathrm{208}}$Pb~\cite{npa424.313} and the potential of  $2n+^{\mathrm{208}}$Pb was substituted by that of $d+^{\mathrm{208}}$Pb~\cite{npa174.485}. Considering the Coulomb interaction exclusively and fixing a normalizing factor 0.27, the theoretical relative energy spectrum is as shown in Fig.~\ref{fig:spec}-b, which agrees well with the experiment result from Labiche et al.~\cite{prl86.600}.

%, where only Coulomb potential was taken into account
%%kinetic calculation
In the following we will figure out how the initial kinetic states of the outgoing particles can be sampled in a Monte Carlo code. In order to generate initial kinetic states of the reaction products, the breakup process is split into two steps, the inelastic scattering of $^{\mathrm{14}}$Be and the breakup of excited $^{\mathrm{14}}$Be$^*$
\begin{eqnarray}%\nonumber
\label{eq:step1}
^{\mathrm{14}} \mathrm{Be} + \mathrm{Pb} &\rightarrow & ^{\mathrm{14}} \mathrm{Be}^*  + \mathrm{Pb}',\\
\label{eq:step2}
^{\mathrm{14}} \mathrm{Be}^* &\rightarrow &^{\mathrm{12}}\mathrm{Be}+ n + n.
\end{eqnarray}
In the first step, the incidence channel is determined according to the experiment setting. On the right side of reaction formula~(\ref{eq:step1}) the mass of $^{\mathrm{14}}$Be$^*$  satisfies $M(^{\mathrm{14}}\mathrm{Be}^*) =E_x + M(^{\mathrm{14}}\mathrm{Be})$, where the excited energy can be written as the sum of the relative energy of the three breakup products and the 2n separation energy of $^{\mathrm{14}}$Be
\begin{equation}
\label{eq:estar}
E_x = E_{rel} + S_{2n}.
\end{equation}
The 2n separation energy $S_{2n}=1.27~(13)~$MeV~\cite{cpc36.1603} is a constant. According to the relative energy spectrum and angular distribution of $^{\mathrm{14}}$Be$^*$ calculated by the CDCC method above, the four dimensional momentum of $^{\mathrm{14}}$Be$^*$ can be generated.

\begin{center}
\includegraphics[width=6cm]{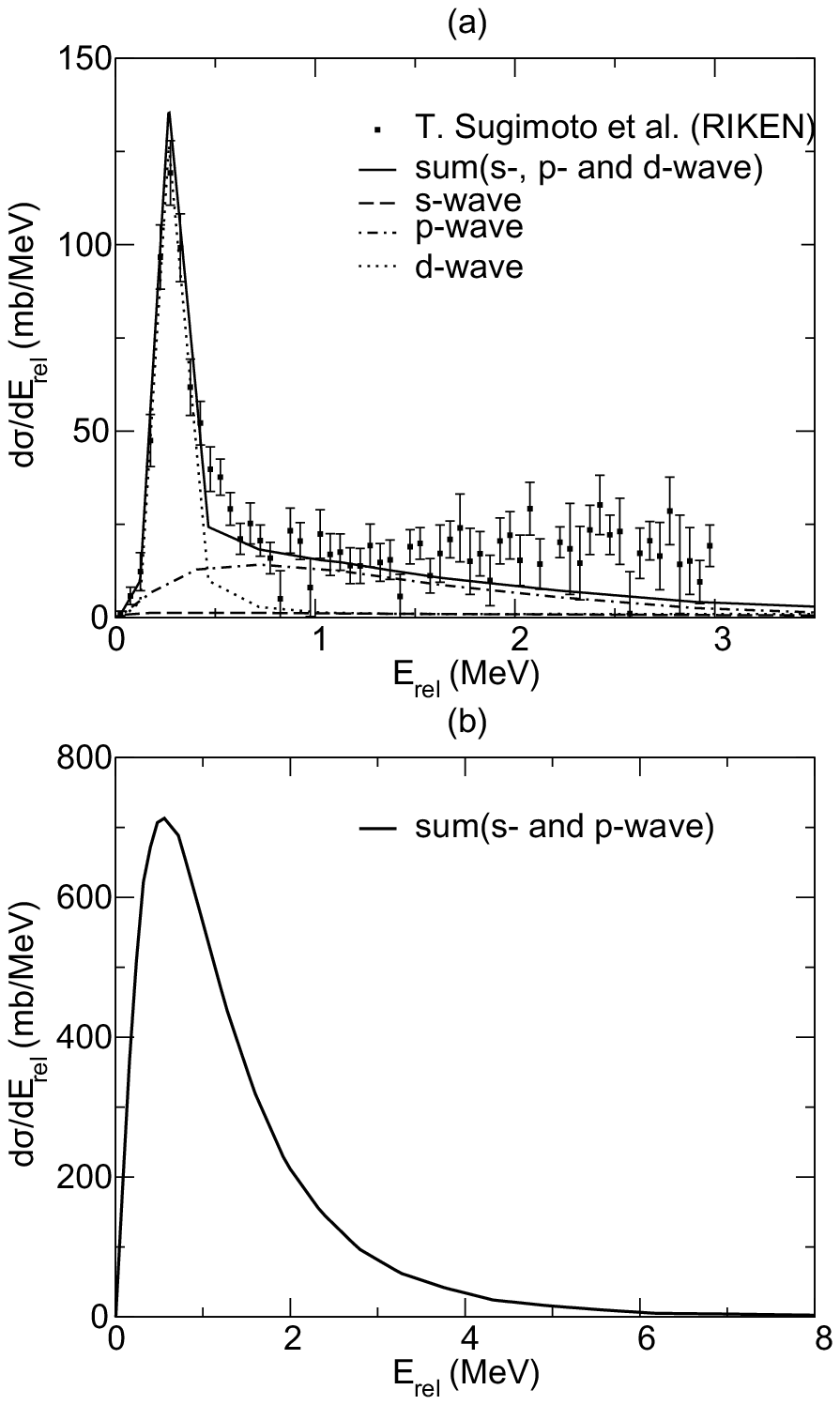}%
\figcaption{\label{fig:spec}(a)The relative energy spectrum calculated by CDCC method and experimental result from \cite{plb654.160} for $^{\mathrm{14}}$Be nuclear breakup on Carbon target. (b)The calculated relative energy spectrum of $^{\mathrm{14}}$Be Coulomb excitation. In the CDCC calculation, the $core+2n$ structure was assumed. }
\end{center}

In order to simplify calculation process the momentum of $^{\mathrm{12}}$Be is determined in the center of mass (c.m.) system of the three outgoing particles($^{\mathrm{12}}\mathrm{Be}+n+n$) and those of the two neutrons are determined in the c.m. system of $2n$.
%The motion of the mass center of the two neutrons relative to the $^{\mathrm{12}}$Be is isotropic in the c.m. system of the products.
In the free phase space the relative kinetic energy taken by $^{\mathrm{12}}$Be is expressed as $E_{rel,^{\mathrm{12}}\mathrm{Be}} = k E_{rel}$, where $k$ is the relative energy partition coefficient varying from 0 to 1. In the outgoing particle c.m. system the differential cross section of the $^{\mathrm{12}}$Be is uniform since there is no direction specific. For the same reason the two neutrons are also isotropic in the c.m. system of $2n$. Thus, the momenta of the $^{\mathrm{12}}$Be and two neutrons are determined in the c.m. system of $^{\mathrm{12}}\mathrm{Be}+2n$ and $2n$ respectively. Consequently through corresponding times of Lorenz transformation the breakup initial states of the products in the laboratory system can be sampled. Assuming $k$ obeys a uniform distribution, we got the neutron angular distribution as shown in Fig.~\ref{fig:nang}, which is in contents with the experimental result of \cite{prl86.600}.

\begin{center}
\includegraphics[width=6cm]{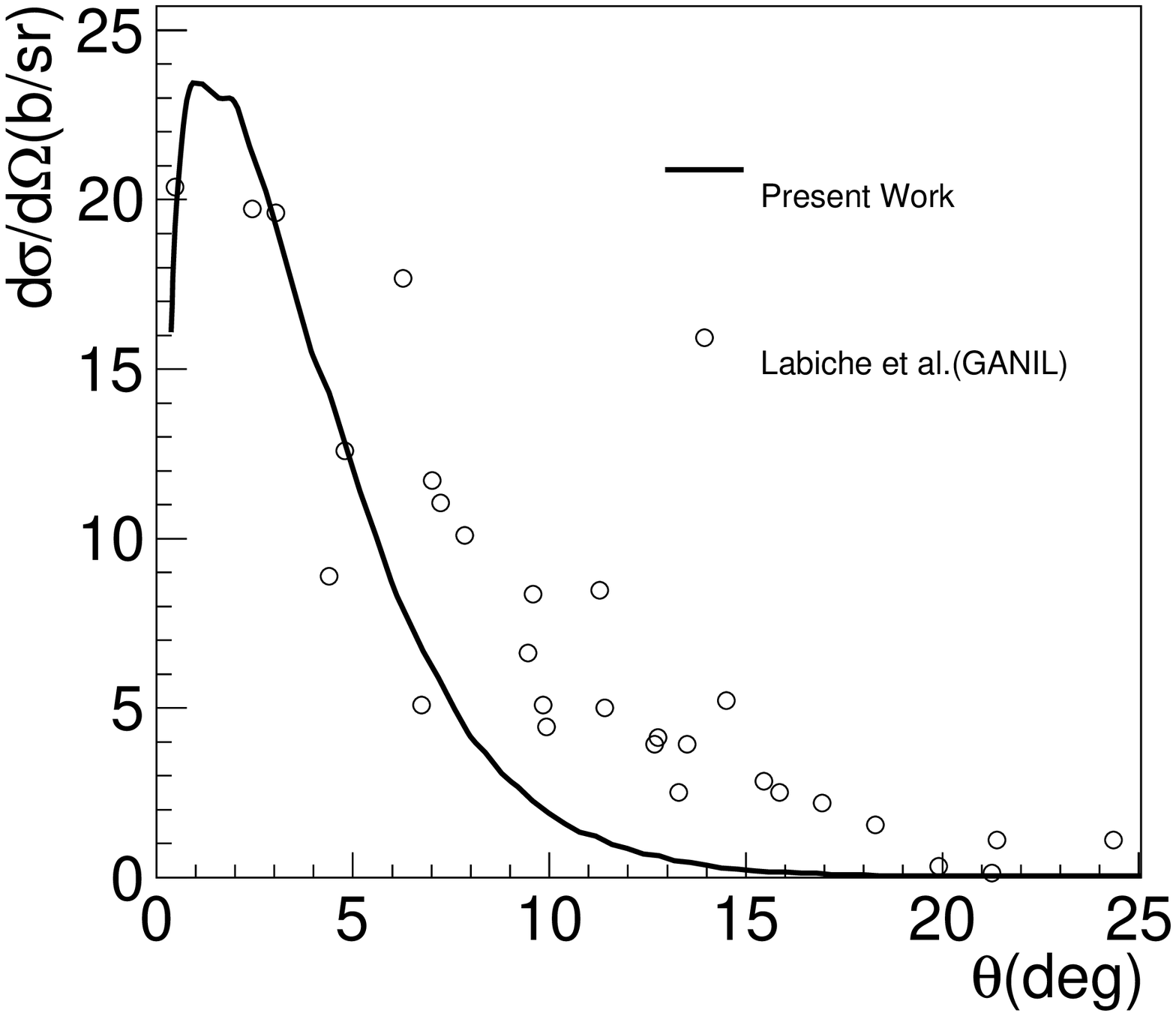}%
\figcaption{\label{fig:nang} Single-neutron angular distribution from the experiment~\cite{prl86.600} (circle) and the simulation of present work(solid line).}
\end{center}

\section{Modeling of the experiment}
Geant4 provides seven major categories of physical processes which may be encapsuled in different modular phy\-sics lists. Some standard physics lists are given for general application in the release package.
In our simulation workspace Geant4 (version 4.10.00) was used and the ``QGSP\_\-BERT\_\-HP'' list was selected as the physics list. In this physics list, the thermal neutron interaction with the scintillator is handled by the data driven high precision neutron (HP) package; a standard electromagnetic (EM) process builder covering ionisation, bremsstrahlung, Coulomb scattering etc. is implemented to deal with EM interactions. To improve the simulation accuracy of slow charged particles, a process builder based on PENELOPE~\cite{nimb254.219} was introduced to replace the standard one. The Coulomb breakup of $^{\mathrm{14}} \mathrm{Be}$ on Pb target as described above was made into a discrete process and added into the physics list ``QGSP\_\-BERT\_\-HP'' according to the physical process constructing principles in Geant4. To improve the computing efficiency of scintillation, we record the electron equivalent energy deposition~\cite{cpc33.860} instead of processing the scintillation photons.

There are convenient and flexible geometric modeling approaches in the Geant4 toolkit. The geometry of the experimental apparatus (without magnetic dipole) for $^{\mathrm{14}}\mathrm{Be}$ Coulomb breakup was modeled as is schematically exhibited in Fig.~\ref{fig:geo}. The double-side silicon strip (SSD) thickness of 1000 micron are chosen and the CsI cross section are 2$\times$2~cm$^2$ (the thickness is set according to the incident energy). They are placed 20~cm and 50~cm behind the target respectively to form a charged ion telescope responsible for outgoing fragments detection. The neutron wall is placed five meters behind the target. It includes five layers of plastic scintillator (BC408) bars with a separation between adjacent layers of 6~cm. The configuration is as described in \cite{cpc14.473}. The cross section of each bar is 6$\times$6~cm$^2$; the number of bars in each layer and the length of the bar are set according to the angular coverage requirement.

\begin{center}
\includegraphics[width=8cm]{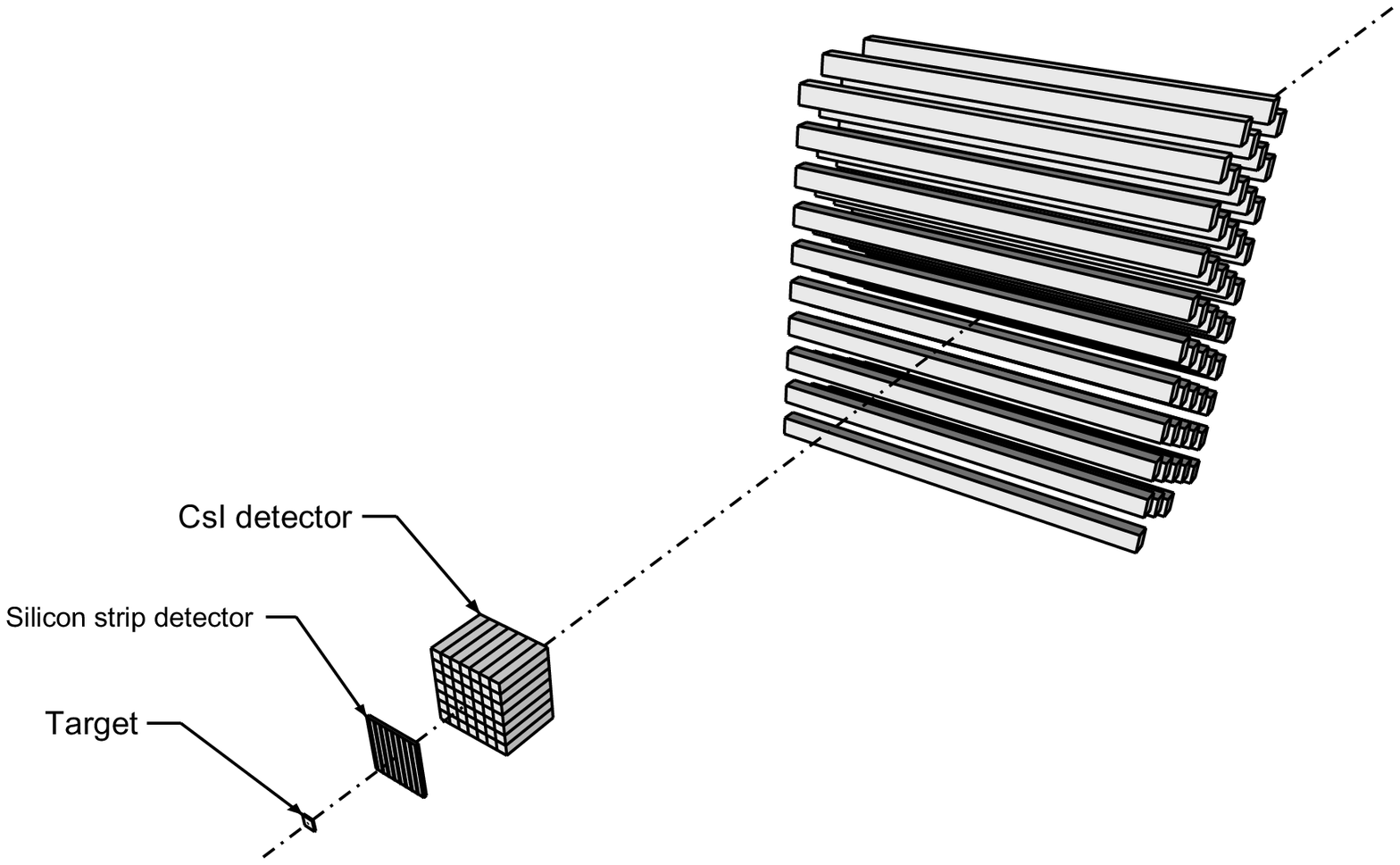}%
\figcaption{\label{fig:geo}The schematic layout of the experimental apparatus for kinematically complete measurement of $^{\mathrm{14}}\mathrm{Be}$ Coulomb breakup. }
\end{center}

\section{Data analysis}
In comparison with the missing mass method the invariant mass method has much higher energy resolution in RIB experiment. Therefore this method is widely implemented in Coulomb breakup reaction experiments of neutron halo nuclei~\cite{prc68.034318,prl96.252502,prl86.600}. Reconstructing the four dimensional momenta of the outgoing particles the invariant mass  $M(^{\mathrm{14}}\mathrm{Be})$ of the excited $^{\mathrm{14}}\mathrm{Be}$ is expressed as
%\end{multicols}
\begin{eqnarray}
\label{eq:invmass}
M(^{\mathrm{14}}\mathrm{Be}^*) &\! = &\! \left[ \left( \sum _i E_i \right)^2-\left( \sum _i \vec{P_i} \right)^2 \right]^{1/2} \nonumber\\
&\! = &\! \left\{ \left[ E(^{\mathrm{12}}\mathrm{Be})+E(n_1)+E(n_2) \right] ^2 \right. \nonumber\\
&&\left. -\left[ \vec{P}(^{\mathrm{12}}\mathrm{Be})+\vec{P}(n_1) + \vec{P}(n_2) \right]^2 \right\}^{1/2},
\end{eqnarray}
%\begin{multicols}{2}
where $E(^{\mathrm{12}}\mathrm{Be})$, $E(n_1)$, $E(n_2)$ are the total energy of the breakup fragment $^{\mathrm{12}}\mathrm{Be}$ and those of two neutrons respectively, and $\vec{P}(^{\mathrm{12}}\mathrm{Be})$, $\vec{P}(n_1)$, $\vec{P}(n_2)$ denote the momenta of three breakup products respectively. The relative energy $E_{rel}$ of the three outgoing particles is extracted from $M(^{\mathrm{14}}\mathrm{Be}^*)$ according to equation~(\ref{eq:estar}).

The study of Coulomb breakup mechanism, correlation of two valent neutrons and the dineutron correlation all depend on the reduced transition probability $B(E1)$ spectroscopy~\cite{prc70.054606,prl96.252502,epja24.63}. According to the semi-classical virtual photon model~\cite{pr163.299} based on first-order perturbation theory, the Coulomb breakup cross section is proportional to the $E1$ virtual photon number $N_{E1}(\theta_{cm}, E_x)$ and $B(E1)$ spectrum as expressed by
\begin{equation}
\label{eq:virtualphoton}
\frac{d^2\sigma}{d\Omega _{cm}dE_{rel}}=\frac{16\pi ^3}{9\hbar c}\frac{dN_{E1}(\theta_{cm}, E_x)}{d\Omega _{cm}}\frac{dB(E1)}{dE_{rel}}.
\end{equation}
 Therefore, the $B(E1)$ spectrum containing nuclear structure can be extracted according to the experimental cross section.

\section{Results and discussion}
The investigation on RIB nuclear physics is based on reverse kinematics and the assumption of thin target. While some experiments~\cite{prl96.252502} have indicated the influence of experimental setup on the spectroscopy investigation. In the following we discuss systematically the effects of the target thickness and the detector performance behind the target on the experiment results.
\subsection{Thin target approximation}
Increasing the target thickness may enhance the reaction yield and make the research of the nuclei close to drip line available. At the same time, it makes the reconstructed relative energy spectrum shape distort. Fixing the incident energy as 35~MeV/nucleon, $^{\mathrm{14}}\mathrm{Be}$ Coulomb breakup on Pb targets was simulated. Ignoring the influence from the performances of the detectors behind the target, the reconstructed relative energy spectra are displayed in Fig.~\ref{fig:tar}-(a). When the target thickness equals 0.1~mm, the relative energy spectrum is quite close to the real. As the target thickness increases, the spectrum peak position move toward high relative energy and the distribution becomes broader. Introducing the mean value $\mu _{rel}$ and standard deviation $\sigma_{rel}$ of the spectrum, the varying tendency depending on target thickness is exhibited in Fig.~\ref{fig:tar}-(b). In this sub graph the values of $\mu _{rel}$ and $\sigma_{rel}$ for target thickness of 0.1~mm are taken as reference, and the vertical coordinates are the ratios of the values for different target thicknesses to the corresponding reference. $\mu _{rel}$ and $\sigma_{rel}$ variate nonlinearly as the target thickness increases. We want to note that $\mu _{rel}$ and $\sigma_{rel}$ increase rapidly when the target thickness is larger than $\sim$0.4~mm, i.e. the reconstructed relative energy spectrum distorts much from that target thickness. In Fig.~\ref{fig:tar}-(c) and Fig.~\ref{fig:tar}-(d) we also present the corresponding $B(E1)$ spectra of different target thickness. The similar variation tendency of $B(E1)$ depending on target thickness is indicated.

In the following we will do some qualitative analysis for the target thickness influence on relative energy spectrum. The effect of the finite target thickness includes two aspects~---~the momentum change of the projectile $^{\mathrm{14}}\mathrm{Be}$ and that of the outgoing fragment $^{\mathrm{12}}\mathrm{Be}$. The target thickness influence on the incident nucleus momentum is usually manifested as the angular distribution change. While it is not the dominant in comparison with that attributed from the behind-target detectors. Hence, we focus on the target effects on the outgoing particles. According to equation~(\ref{eq:estar}), the relative energy spectrum only depends on the experiment outcome variable $M(^{\mathrm{14}}\mathrm{Be}^*)$, which can be obtained by  equation~(\ref{eq:invmass}). Squaring equation~(\ref{eq:invmass}), we get the mass squared $M^2(^{\mathrm{14}}\mathrm{Be}^*)$ of $^{\mathrm{14}}\mathrm{Be}^*$ expression as following
\begin{equation}
\label{eq:mep}
\begin{split}
& M^2(^{\mathrm{14}}\mathrm{Be}^*) = \sum _{i=1}^{3} M_i^2 -2\left[ P(n_1)P(n_2)cos\theta _1 -E(n_1)E(n_2)\right] \\
& \qquad\qquad -2\left[ P(n_1)P(^{\mathrm{12}}\mathrm{Be})cos\theta _2 +P(n_2)P(^{\mathrm{12}}\mathrm{Be})cos\theta _3 \right], \\
& \qquad\qquad +2\left[ E(n_1) + E(n_2)\right]\left[ M^2(^{\mathrm{12}}\mathrm{Be}) + P^2(^{\mathrm{12}}\mathrm{Be})\right]^{1/2}, \\
\end{split}
\end{equation}
where $\sum _{i=1}^{3} M_i^2$ is the sum of the mass squared of the three outgoing particles; $\theta _1$, $\theta_2$ and $\theta _3$ are the angles between two neutrons as well as between two neutrons and $^{\mathrm{12}}\mathrm{Be}$ respectively. Equation~(\ref{eq:mep}) indicates that the target affects the relative energy by changing $^{\mathrm{12}}\mathrm{Be}$ momentum of $\vec{P}(^{\mathrm{12}}\mathrm{Be})$, which may split into the momentum value $P(^{\mathrm{12}}\mathrm{Be})$ and the directions $\theta_2$ and $\theta _3$. The changing of $P(^{\mathrm{12}}\mathrm{Be})$ after it is produced in the target depends on the energy deposition of $^{\mathrm{12}}\mathrm{Be}$ in the target, which could be briefly analyzed by the energy deposition theory of ions. According to the Bethe-Block formula there is approximately $\Delta E_{k} \propto \Delta x z^2/(E_{k}/A)$, where $\Delta x$ is the medium thickness; $z$ and $E_{k}/A$ are atomic number and the energy per nucleon of a incident nucleus respectively. With the type of projectile and incident energy fixed, the increase of target thickness $\Delta x$ will make the energy difference increase linearly. At the same time, the scattering angle becomes larger according to Rutherford scattering theory. That is why the parameters indicating the spectrum distortion increase with the target thickness non-linearly. By simulation we also found that the experiment with higher energy per nucleon may tolerate thicker target, which is also consistent the qualitative analysis based on the Bethe-Block formula.

\begin{center}
\includegraphics[width=9cm]{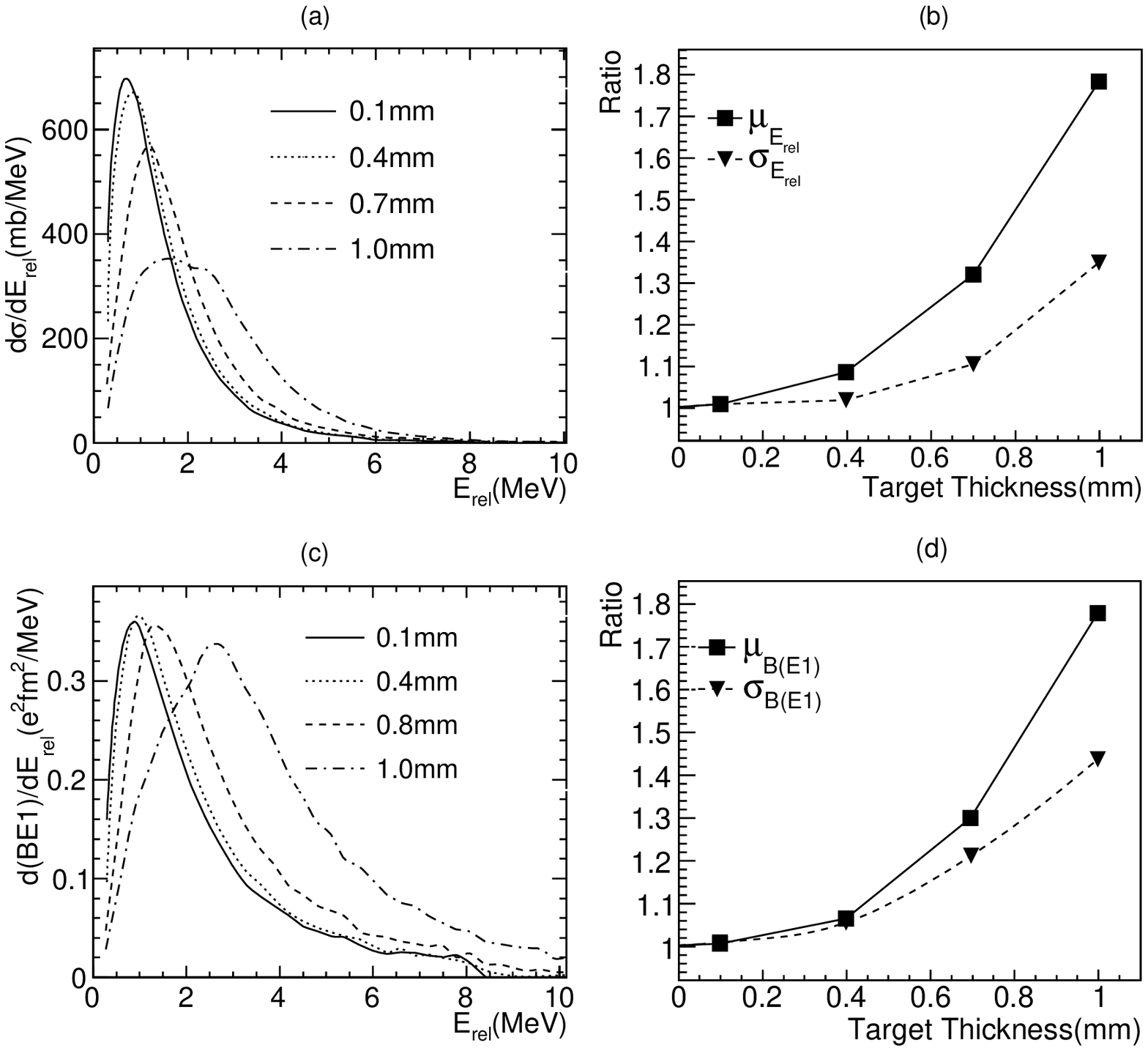}%
\figcaption{\label{fig:tar} (a)The reconstructed relative energy spectra for different target thickness with invariant mass method. (b) displays how the target thickness influence the mean value $\mu _{rel}$ and standard deviation $\sigma_{rel}$ of the relative energy spectrum. (c) and (d) indicate how the target thickness influences the $B(E1)$ spectrum. In subgraph (b) and (d), $\mu$ and $\sigma$ with the target thickness of 0.1~mm was selected as the reference, and the vertical coordinates are the ratios of the values of different target thicknesses to the corresponding reference values.}
\end{center}

\subsection{Neutron detector performances}
According to the simulation, the performance of the charged ion telescope today does not have significant influence on the spectroscopic study of the Coulomb breakup of $^{\mathrm{14}}\mathrm{Be}$ on a Pb target. So we focus on the neutron wall performance effects, which has also been interpreted in the $^{\mathrm{11}}\mathrm{Li}$ experimental investigation~\cite{prl96.252502}. If the efficiency for events with certain relative energy is extremely low, there will be a poor statistic in experiment, which may cause incorrect understanding of the nuclear structure. For different experimental setting Fig.~\ref{fig:neff} exhibits the calculated efficiency curves of neutron walls, where the problem of cross talk~\cite{cpc14.473} is not considered. The efficiency varies with the relative energy as is shown by the solid line when the angular acceptance is $4^{\circ}$~(half angle) in x and y direction and the incident energy of $^{\mathrm{14}}\mathrm{Be}$ is 35~MeV/nucleon. It is quite similar as that of the neutron wall of MSU used in the experiment~\cite{prc48.118} with a very low efficiency especially for high relative energy. Keeping the same experimental setting but doubling the angular coverage as $8^{\circ}$, there is an apparent rise for the efficiency as shown by dashed line. For different incident energy per nucleon a neutron wall also manifests different characteristics of efficiency. The higher the incident energy the more forward the outgoing neutrons, which will enhance the efficiency of a neutron wall with certain angular acceptance. At the same time, there will be more events with two neutrons hitting on the same scintillator bar. Hence, the efficiency drops at low relative energy. It is indicated in Fig.~\ref{fig:neff}-b with an incident energy of 280~MeV/nucleon. The regular fluctuation of the efficiency curve are due to the incompact configuration of the scintillator bars.

\begin{center}
\includegraphics[width=9cm]{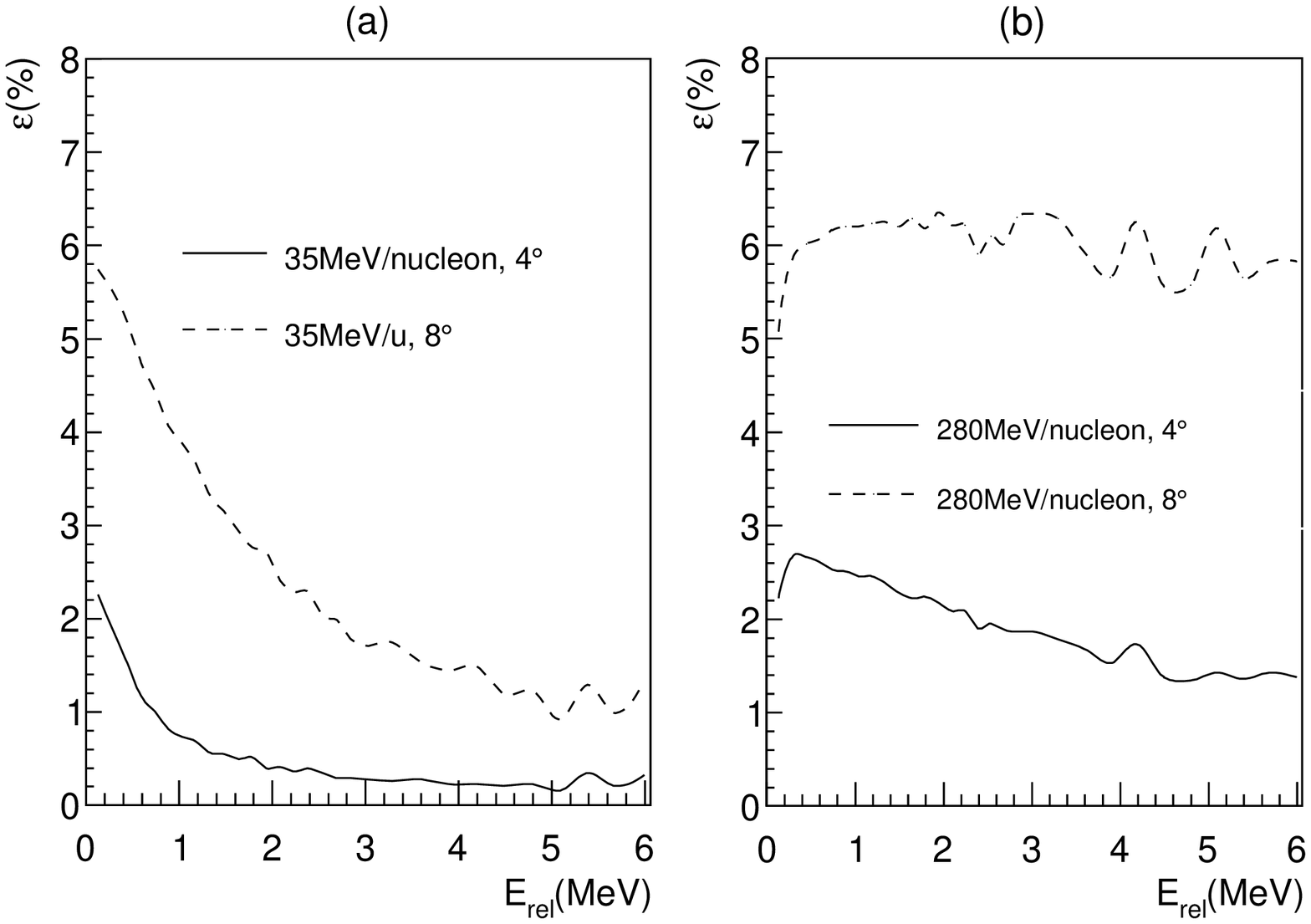}%
\figcaption{\label{fig:neff} The neutron wall efficiency ($\varepsilon$) curves for different experiment design of $^{\mathrm{14}}\mathrm{Be}$+Pb given by the simulation. The incident energy of $^{\mathrm{14}}\mathrm{Be}$ is 35MeV/nucleon in (a) and 280MeV/nucleon in (b). }
\end{center}

\section{Conclusions}
A two body model ($^{\mathrm{12}}\mathrm{Be}+$dineutron) was implemented to describe the 2n halo nucleus $^{\mathrm{14}}\mathrm{Be}$. Hence, the three-body CDCC calculation was implemented successfully in the $^{\mathrm{14}}\mathrm{Be}$ breakup reactions at Carbon target and lead target respectively.
%Fixing the s-wave spectroscopic factor of the dineutron as 0.27 the relative energy spectrum calculated by CDCC method agrees with the experimental result from \cite{plb654.160} pretty well. With the same structure parameters the Coulomb breakup reaction of $^{\mathrm{14}}\mathrm{Be}$ was treated with CDCC calculation. The calculated relative energy and single neutron differential cross section are also in consent with the exist experimental results~\cite{prl86.600}.
It is interpreted that the three-body CDCC calculation is fit for the description of the breakup of 2n halo nuclei in the given scenarios. Combining the theoretical calculation with Geant4 toolkit a simulation workspace for the kinematically complete measurement of $^{\mathrm{14}}\mathrm{Be}$ Coulomb breakup was developed. By the simulation it was found that the relative energy spectrum is very sensitive to the target thickness in the heavy target Coulomb excitation experiment.
%According the existing experimental data, the Coulomb excitation of neutron halo nuclei is a kind of soft excitation. Its relative energy spectra peak at low relative energy.
As for the detectors behind the target, the neutron wall substantially affects the uncertainty of the energy spectroscopy. Thus, it was discussed how the angular acceptance and the incident energy influence the neutron wall efficiency.
%The granularity of the neutron wall is another critical influence factor besides the angular acceptance of the whole neutron wall.
The corresponding results are also referential for Coulomb breakup research of other neutron halo nuclei.\\

\acknowledgments{We thank the collaborators from Peking University and Institute of Modern Physics for their helpful suggestions. Thanks also goes to Dr. Pang and Dr. Moro for their help in FRESCO usage. }

\end{multicols}
\vspace{-1mm}
\centerline{\rule{80mm}{0.1pt}}
\vspace{2mm}

\begin{multicols}{2}

\end{multicols}

\clearpage

%\end{CJK*}
\end{document}